\documentclass[aps,prl,twocolumn,showpacs,amsmath,amssymb,superscriptaddress]{revtex4-1}
\usepackage{graphicx}
\usepackage{dcolumn}
\usepackage{bm}
\usepackage{epsfig}
\usepackage{epstopdf}
\usepackage{float}	
\usepackage{color}
\usepackage[title]{appendix}
\usepackage{mathrsfs}
\usepackage{enumitem}\usepackage{hyperref}
\usepackage{braket}\usepackage{mathtools}
\usepackage{amsmath}\usepackage{ulem}

\usepackage{tabularx}
\usepackage{float}
\newcolumntype{C}{>{\centering\arraybackslash}X}

\begin{document}
\title{Anomalous Long-Distance Coherence in Critically-Driven Cavity Magnonics}

\author{Ying~Yang}
\affiliation{Department of Physics and Astronomy, University of Manitoba, Winnipeg, Canada R3T 2N2}

\author{Jiguang~Yao}
\affiliation{Department of Physics and Astronomy, University of Manitoba, Winnipeg, Canada R3T 2N2}

\author{Yang~Xiao}
\affiliation{Department of Physics, Nanjing University of Aeronautics and Astronautics, Nanjing 210016, China}

\author{Pak-Tik~Fong}
\affiliation{Department of Physics, Simon Fraser University, Burnaby, Canada V5A 1S6}

\author{Hoi-Kwan~Lau}
\affiliation{Department of Physics, Simon Fraser University, Burnaby, Canada V5A 1S6}
\affiliation{Quantum Algorithms Institute, Surrey, British Columbia, Canada V3T 5X3}

\author{C.-M.~Hu} \email{hu@physics.umanitoba.ca; \\URL: http://www.physics.umanitoba.ca/$\sim$hu}
\affiliation{Department of Physics and Astronomy, University of Manitoba, Winnipeg, Canada R3T 2N2}

\date{\today}

\begin{abstract}
Developing quantum networks necessitates coherently connecting distant systems via remote strong coupling. Here, we demonstrate long-distance coherence in cavity magnonics operating in the linear regime. By locally setting the cavity near critical coupling with travelling photons, non-local magnon-photon coherence is established via strong coupling over a 2-meter distance. We observe two anomalies in this long-distance coherence: first, the coupling strength oscillates twice the period of conventional photon-mediated couplings; second, clear mode splitting is observed within the cavity linewidth. Both effects cannot be explained by conventional coupled-mode theory, which reveal the tip of an iceberg of photon-mediated coupling in systems under critical driving. Our work shows the potential of using critical phenomena for harnessing long-distance coherence in distributed systems.

\end{abstract}

\maketitle

\textit{Introduction.--}
Coherent dynamics enabled by strong light-matter interactions \cite{chang2018colloquium,rameshti2022cavity,chumak2022advances} contributes significantly to the advancements in quantum science and technology, which are typically achieved by overlapping electromagnetic fields. When systems are separated by macroscopic distances, the direct coupling is hindered due to the reduced overlap of fields, posing a challenge for establishing and preserving long-distance coherence.

To achieve long-distance coherence, several methods are employed, such as optomechanical systems \cite{xuereb2012strong}, superconducting cavities \cite{li2022coherent}, topological edge states \cite{hetenyi2022long} and also surface acoustic waves \cite{schutz2017universal}.
One of the most interested methods involves travelling photons in microwave waveguides, laser beams, or optical fibers \cite{yanik2004time,xu2006experimental,luo2021nonlocal,mukhopadhyay2022anti}. Indirect couplings between distant resonators are generated through their cooperation of dissipations to travelling photons \cite{metelmann2015nonreciprocal,wang2019nonreciprocity}, and have garnered broad interest in quantum optics \cite{lodahl2017chiral,foster2019tunable} and waveguide quantum electrodynamics \cite{sheremet2023waveguide,mirhosseini2019cavity,corzo2019waveguide}.

The main difficulty for implementing this approach is dissipation: in conventional systems where a pair of resonators are connected by travelling photons, the same photons that mediate the coupling induce an extrinsic dissipation, causing photon-induced decoherence that erases the photon-mediated coherence \cite{fan2003temporal,lalumiere2013input,buchmann2015quantum,karg2019remote}. In order to establish long-distance coherence, special methods are employed for either enhancing the coupling or suppressing the dissipation. For example, long-distance strong coupling has been demonstrated by terminating the waveguide with mirrors \cite{majer2007coupling,wen2019large,lin2019scalable,mirhosseini2019cavity}, adopting a light loop \cite{karg2019remote,karg2020light}, constructing resonators as giant atoms \cite{kockum2018decoherence,kannan2020waveguide,wang2022giant}, or utilizing a gain-driven cavity \cite{yao2023coherent,rao2023meterscale}. While these methods enhance the cooperativity, they also come with drawbacks such as bandwidth limitations, stability constraints, design challenges, and nonlinear disruptions.

Recently, cavity magnonics \cite{rameshti2022cavity} has emerged as a versatile platform for engineering light-matter interactions. Among many advantages, it enables the incorporation of critical phenomena, such as exceptional points and bound states in the continuum in coupled systems. Experiments have found that by using critical phenomena, dissipations can be harnessed as a resource \cite{yang2020unconventional}. This sparks curiosity about whether cavity magnonics could pave the way for establishing coherence over long distances by utilizing critical phenomena.

This work experimentally explores such a frontier by studying critically-driven cavity magnonics. A cylindrical dielectric cavity \cite{petosa2007dielectric} is used to study its remote coupling with an yttrium iron garnet (YIG) sphere \cite{serga2010yig}. The coupling is mediated by photons travelling in coaxial cables over 2 meters long. Near the critical-driving condition where the cavity is critically coupled to travelling photons, we observe a normal mode splitting that demonstrates coherent energy exchange mediated by photons travelling over a long distance. Moreover, by comparing with standard theories for photon-mediated coupling, we show that the observed coupling is anomalous in both the coupling strength and its dependence on the phase of the travelling photons.

\begin{figure} [!t]
\begin{center}
\includegraphics[width=8.7 cm]{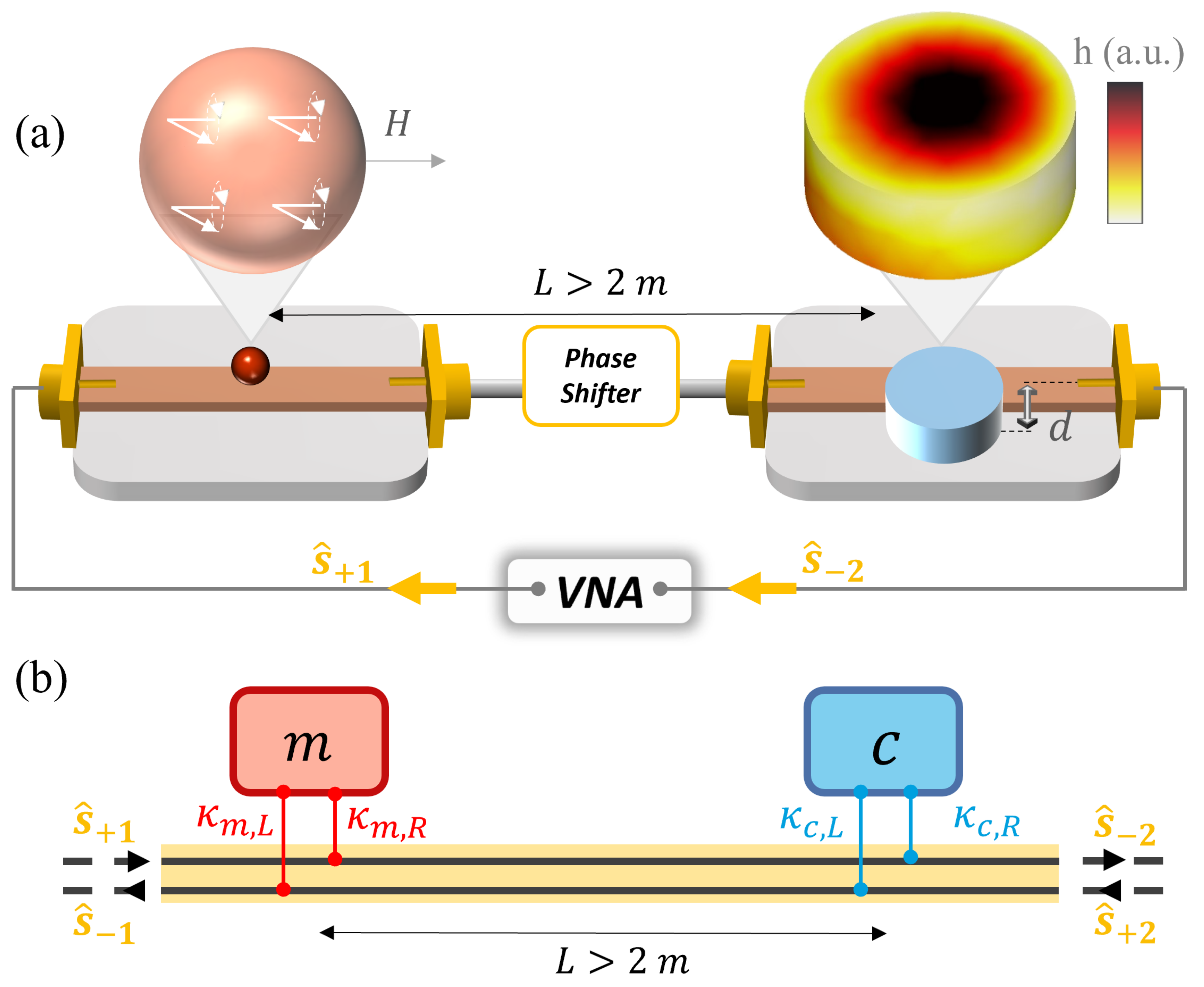}
\caption{\textbf{Setup.} (a) The YIG sphere is placed on a microstrip. An external magnetic field $H$ applied parallel to the microstrip drives the magnon mode with the uniform precession. The dielectric cylinder is placed on another microstrip. The cavity mode profile in the top right resembles the TE01 mode. The lateral spacing $d$ between the centers of the cavity and the microstrip is tunable by using a step motor. The magnon and cavity modes are coupled remotely via photons travelling in coaxial cables that connect the microstrips. The travelling phase $\Phi = \Phi_L + \Delta\Phi$ is controlled by an inserted phase shifter that tunes $\Delta\Phi$, while $\Phi_L = 2\pi L/ \lambda$ is set by the photon propagation distance $L$. The transmission spectra $S_{21}(\omega)$ are measured by connecting the end ports of the microstrips to the Vector Network Analyzer.
(b) Theoretical model. The magnon and cavity photon modes are coupled to the travelling photons with rates $\kappa_{m,L(R)}$ and $\kappa_{c,L(R)}$, respectively, and measured by input $\hat{s}_{+1}$ and output $\hat{s}_{-2}$ fields. }
\label{Fig1}
\end{center}
\end{figure}

\textit{Setup.--}
Our setup is shown in Fig. \ref{Fig1}(a). The YIG sphere with 1 mm diameter is placed on a $4.65$ mm-wide microstrip. It is biased by an external magnetic field $H$ applied parallel to the microstrip, which controls the resonant frequency $\omega_m = \gamma_e \mu_0 (H+H_A)$ of the magnon mode. Here, $\mu_0$ is the vacuum permeability. From calibrations \cite{Supp}, we get the gyromagnetic ratio $\gamma_e = 2\pi \times 22.4$ GHz/T, the anisotropy field $\mu_0 H_A = -7.1$ mT, the intrinsic magnon damping rate $\alpha_0/2\pi$ = 0.8 MHz, and the rates of extrinsic damping of the magnon mode to the left-going ($\kappa_{m,L}/2\pi$ = 8 MHz) and right-going ($\kappa_{m,R}/2\pi$ = 7 MHz) travelling waves (schematically shown in Fig. \ref{Fig1}(b)).

 A dielectric cylinder \cite{petosa2007dielectric}, with a diameter of $9.1$ mm, a height of $3.7$ mm and a dielectric constant of $34$, is placed on another $4.65$ mm-wide microstrip. In the top right of Fig. \ref{Fig1}, the simulated cavity mode profile is plotted, resembling the TE01 mode where a magnetic dipole is along the vertical axis of the cylinder. The lateral spacing $d$ between the centers of the cavity and the microstrip is tunable by a step motor. The two microstrips are connected by coaxial cables with a total length $L$, and a phase shifter is inserted into the cables.

Our goal is to establish non-local coherence between the magnon and cavity modes via photons travelling in the coaxial cables. At $\omega/2\pi = 6.2$ GHz (wavelength $\lambda$ = 32.7 mm), the travelling phase between the magnon and photon modes is $\Phi = \Phi_L + \Delta\Phi$, where $\Phi_L = 2\pi L/ \lambda >$ 128$\pi$ is the phase of the photons propagating over a distance of $L >$ 2 m \cite{Supp}. $\Delta\Phi$ is controlled by the phase shifter, precisely tuning $\Phi$ over a period of $2\pi$. The transmission spectra $S_{21}(\omega)$ are measured by connecting the end ports of the microstrips to the Vector Network Analyzer. The experiments are performed in the linear dynamics regime by setting the input power at -10 dBm.

\textit{Critical coupling.--}
We calibrate the critical coupling \cite{pozar2011microwave}  between the cavity and microstrip to construct critically-driven cavity magnonics. The condition is set by the lateral spacing $d$ that controls the cavity-microstrip coupling. In the calibration measurement, the YIG sphere is unbiased by setting $H$ = 0, and the phase shifter is set at $\Delta\Phi = \pi$.

\begin{figure} [!t]
\begin{center}
\includegraphics[width=8.8 cm]{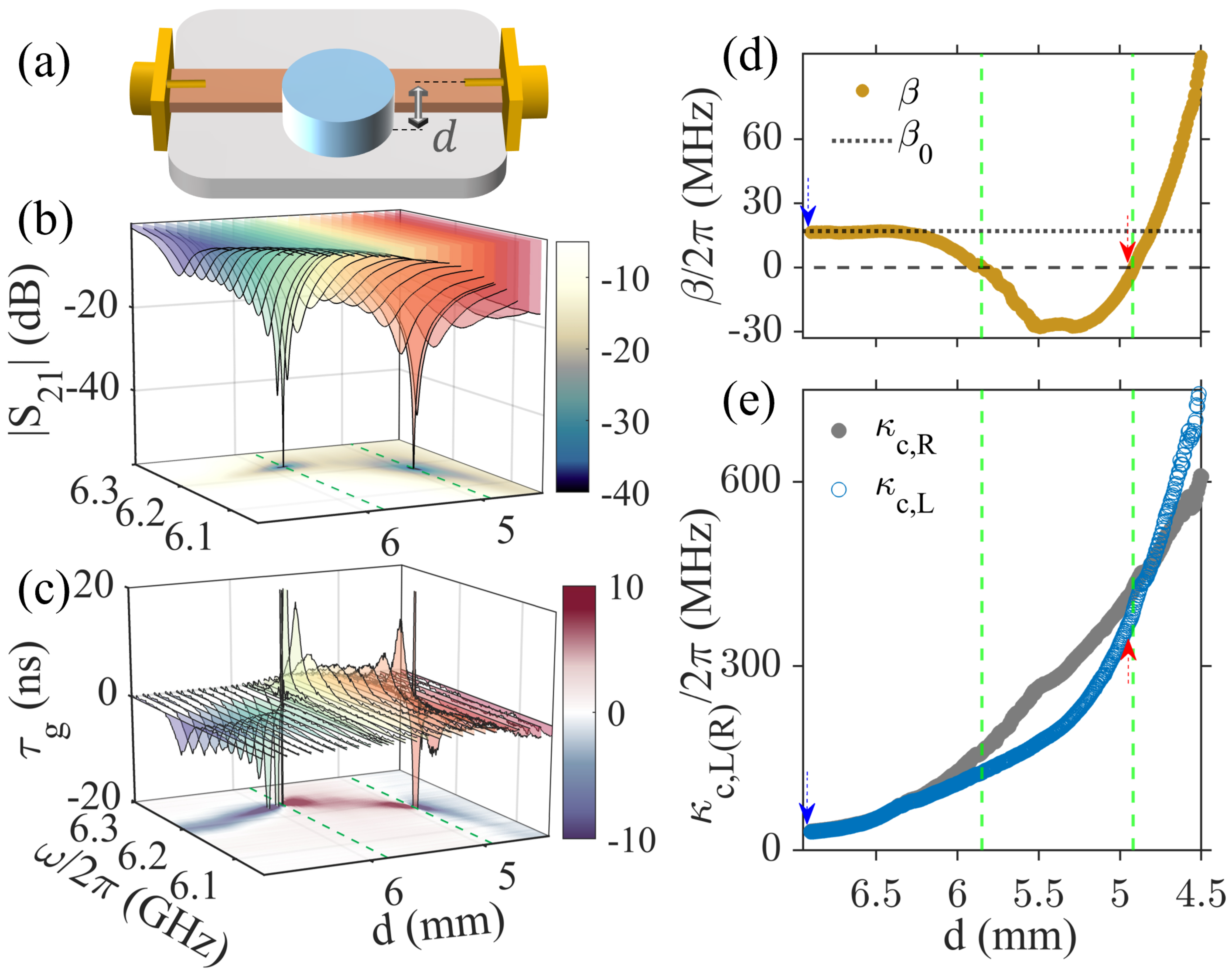}
\caption{\textbf{Critical coupling.} (a) The dielectric cylinder is calibrated by tuning its lateral spacing $d$ to the microstrip. (b) The transmission amplitude $|S_{21}|$ and (c) group delay $\tau_g$ reveal two critical coupling conditions marked by the green dashed lines. (d) The effective damping rate $\beta/2\pi$ and (e) the extrinsic damping rate $\kappa_{c, L(R)}/2\pi$ are extracted by fitting the transmission using Eq. \ref{eq:s21cavity}. The dotted line in (d) shows the intrinsic damping rate $\beta_0/2\pi$. The arrows in (d) and (e) mark the cavity setting conditions for the data presented in Figs. \ref{Fig3} and \ref{Fig4}.}
\label{Fig2}
\end{center}
\end{figure}

Figs. \ref{Fig2}(b) and (c) plot the measured transmission amplitude $|S_{21}(\omega)|$ and the group delay $\tau_g(\omega) = -\partial\angle S_{21}/\partial \omega$ of the cavity, respectively, where $\angle S_{21}$ is the transmission phase. By changing $d$, the cavity mode frequency $\omega_c$ shifts non-monotonically. Two sharp dips are observed at $d$ = 4.90 and 5.80 mm in Fig. \ref{Fig2}(b), each marked by a green dashed line. These are the critical coupling conditions, as we explain below. At these conditions, $|S_{21}(\omega_c)|$ approaches zero in the linear scale, while $\tau_g (\omega_c)$ switches abruptly between positive and negative infinities. In the region between the two dashed lines, $\tau_g (\omega_c) > 0$; outside of the region, $\tau_g (\omega_c) < 0$.

Using the input-output theory \cite{gardiner1985input,yu2020chiral}, we derive \cite{Supp},
\begin{equation}
\begin{split}
S_{21} = \frac{\omega-\omega_c + i(\beta_0+\kappa_{c,L}/2-\kappa_{c,R}/2)}{\omega-\omega_c + i(\beta_0+\kappa_{c,L}/2+\kappa_{c,R}/2)},
\end{split}
\label{eq:s21cavity}
\end{equation}
where the intrinsic damping rate of the cavity mode is calibrated as $\beta_0/2\pi$ = 17 MHz. $\kappa_{c,L}/2\pi$ and $\kappa_{c,R}/2\pi$ are the rates of extrinsic damping of the cavity mode to the left-going and right-going travelling waves, respectively. Both of them depend on  $d$.

Eq. \ref{eq:s21cavity} shows that $1/|S_{21} (\omega)|$ has a Lorentzian line shape with HWHM = $|\beta|$, where $\beta$ is an effective damping rate defined as
\begin{equation}
\beta=\beta_0+\kappa_{c,L}/2-\kappa_{c,R}/2.
\label{eq:damping}
\end{equation}
It models the cavity as a loaded oscillator with $\tilde{\omega}_c =\omega_c-i\beta$, where the loading effect of the travelling photons impacts both HWHM and $\tau_g$. At the resonance, $\tau_g (\omega_c) \propto - \frac{1}{\beta}$, so that $\beta$ determines the sign of $\tau_g (\omega_c)$: when $\beta >$ 0, we have $\tau_g (\omega_c) < 0$, the cavity is under coupled to the travelling photons; when $\beta <$ 0, the cavity is over coupled, leading to $\tau_g (\omega_c) > 0$; at $\beta = 0$, the cavity is critically coupled, which is a singularity where $\tau_g (\omega_c)$ approaches infinity \cite{pozar2011microwave}.

Using Eq. \ref{eq:s21cavity} to fit the measured $S_{21}$ spectra \cite{Supp}, we deduce the fitting parameters $\kappa_{c,L}$ and $\kappa_{c,R}$ and plot them in Fig. \ref{Fig2}(e). $\beta$ is determined from Eq. \ref{eq:damping} and plotted in Fig. \ref{Fig2}(d). Note that due to an interference effect \cite{kockum2018decoherence,kannan2020waveguide,bo2017controllable,chen2022nonreciprocal,1075416,Huang1994,ghulinyan2013oscillatory}, when $d$ decreases, $\kappa_{c,L}$ and $\kappa_{c,R}$ increase differently with the enhanced field overlapping of the cavity and microstrip, so that $\beta$ oscillates which leads to two critical coupling conditions at $d$ = 4.90 and 5.80 mm \cite{Supp}. Near both critical conditions, the cavity mode dynamically functions as a loaded high-Q
oscillator with nearly zero effective damping, which is extremely sensitive to detect long-distance coherence, as we demonstrate below in two experiments.

\textit{Photon-mediated long-distance coupling.--}
The 1st experiment, performed at different $\Phi$ while setting the $H$-field at $\omega_m$ = $\omega_c$, searches for the evidence of mode splitting caused by photon-mediated coupling. Typical results are comparatively presented for $d$ = 4.92 mm [marked by the red arrow in Fig. \ref{Fig2}(d)] and $d$ = 6.90 mm (blue arrow). Table \ref{Table} lists the cavity parameters at these settings.

\begin{figure} [!t]
\begin{center}
\includegraphics[width=9 cm]{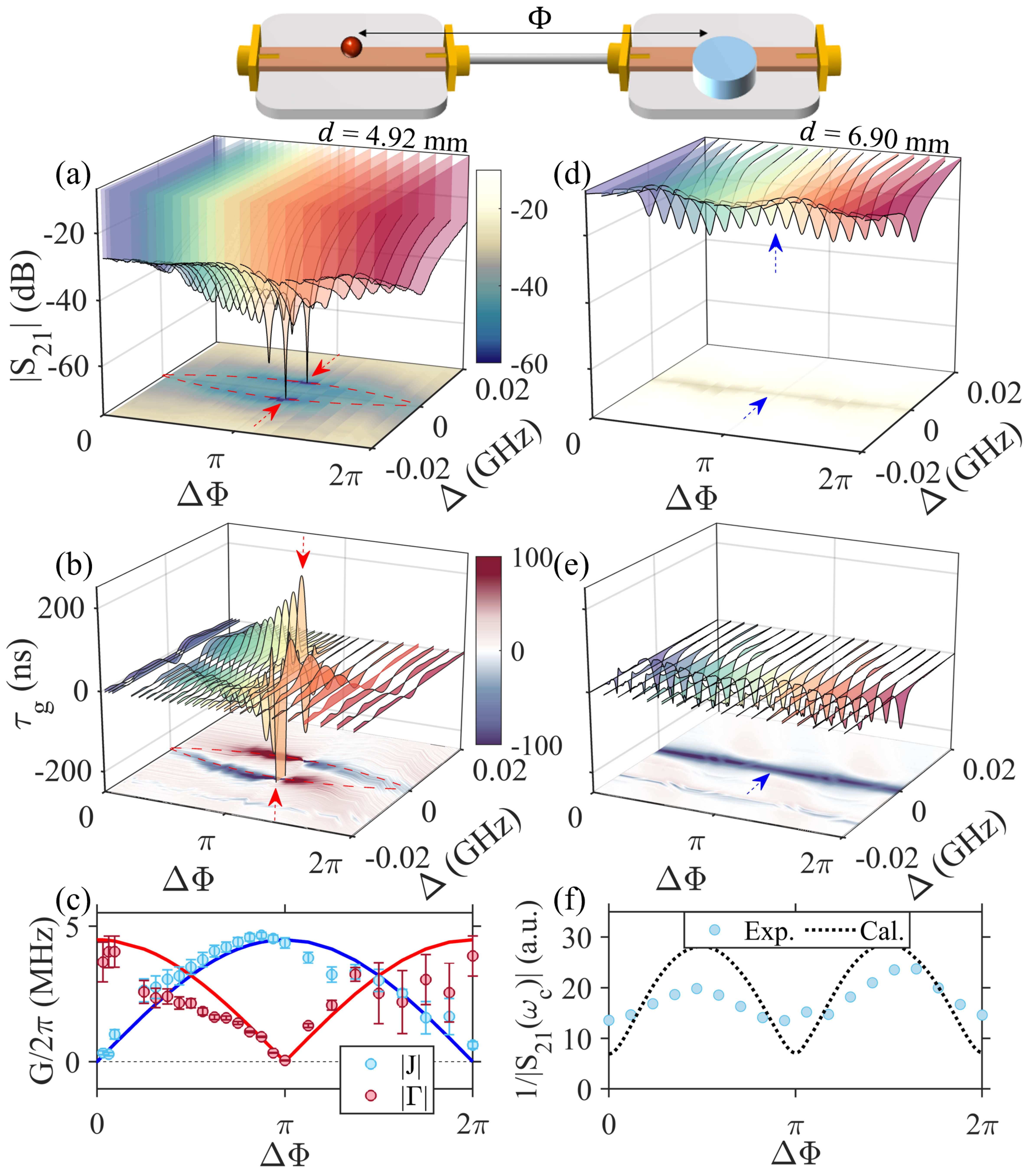}
\caption{\textbf{Long-distance coupling: the phase dependency.} The left and right panels compare the results measured at $d$ = 4.92 and 6.90 mm, respectively. (a) $\&$ (d) Transmission amplitude $|S_{21}|$, (b) $\&$ (e) group delay $\tau_g$ measured as a function of the frequency detuning $\Delta = (\omega-\omega_c)/2\pi$ at different phase setting $\Delta \Phi$. The dashed curves in (a) and (b) are calculated by using Eq. \ref{eq:s21} with $\eta = 2$ and $\delta = 0.996$. (c) The real and imaginary components of the complex coupling strength $G = J+i\Gamma$ are plotted as blue and red circles, while the solid sinusoidal curves are added to guide the eye. (f) The measured inverse amplitude $1/|S_{21}(\omega_c)|$ are plotted in comparison with the result calculated using Eq. \ref{eq:s21} with $\eta = 1$ and $\delta = 1$.}
\label{Fig3}
\end{center}
\end{figure}

\begin{table}[H]
\centering
\caption{Cavity parameters at typical settings near or away from the critical coupling (c.c), calibrated at $\Delta\Phi$ = $\pi$.}
\begin{tabularx}{\columnwidth}{CCC}
\hline\hline
Cavity Setting & near c.c. & away from c.c. \\
\hline
\(d\) (mm) & 4.92 & 6.90 \\
\(\omega_c/2\pi\) (GHz) & 6.181 & 6.203 \\
\(\kappa_{c,L}/2\pi\) (MHz) & 332.4 & 37.0 \\
\(\kappa_{c,R}/2\pi\) (MHz) & 370.0 & 37.0 \\
\(\beta/2\pi\) (MHz) & -1.8 & 17.0 \\
\hline\hline
\end{tabularx}
\label{Table}
\end{table}

At $d$ = 4.92 mm, three key features are observed: (i) mode splitting is found in $|S_{21}|$ plotted in Fig. \ref{Fig3}(a), which depends on the phase $\Phi = \Phi_L + \Delta\Phi$ (without loss of generality, we denote $\Delta\Phi$ = $\pi$ at the maximum mode splitting). This feature demonstrates that the magnon and cavity photon modes are coupled remotely by the travelling photons. Such a non-local hybridization leads to two normal modes $\tilde{\omega}_{\pm} = \omega_{\pm} - i\delta_{\pm}$, where $\omega_{\pm}$ and $\delta_{\pm}$ are the resonant frequencies and damping rates, respectively. (ii) the measured group delay of the hybridized modes exhibits antisymmetric phase dependence as shown in Fig. \ref{Fig3}(b). For $\Delta\Phi < \pi$, we observe $\tau_g > 0$ for the higher-frequency mode $\tilde{\omega}_{+}$, and $\tau_g < 0$ for $\tilde{\omega}_{-}$; when $\Delta\Phi > \pi$, the results are reversed. This feature is associated with the anomalous phase period that we will show below. (iii) for both modes, $\tau_g$ switches abruptly between positive and negative infinities at $\Delta\Phi \simeq \pi$, where the non-locally hybridized modes are critically coupled with the travelling photons.

As the first step for understanding these intriguing features, we model the non-local hybridization as two harmonic oscillators ($\tilde{\omega}_m =\omega_m-i\alpha$ and $\tilde{\omega}_c =\omega_c-i\beta$) coupled by a complex rate $G$. Here, $\alpha \equiv \alpha_0+\kappa_{m,L}/2-\kappa_{m,R}/2$ = 1.3 MHz is the effective damping rate for the magnon mode. Fitting $1/|S_{21}|$ with the lineshapes involving two resonances, we determine $\tilde{\omega}_{\pm} = \omega_{\pm} - i\delta_{\pm}$, from which we extract \cite{Supp},
\begin{equation}
G = J+i\Gamma  = \sqrt{(\tilde{\omega}_+ -\tilde{\omega}_-)^2-(\tilde{\omega}_c -\tilde{\omega}_m)^2}.
\label{eq:g}
\end{equation}
Fig. \ref{Fig3}(c) plots the phase dependence of the real ($|J|$) and imaginary ($|\Gamma|$) components of $G$. Both $J$ and  $\Gamma$ follow a sinusoidal solid curve, and $G$ changes from purely real (coherent coupling) to purely imaginary values (dissipative coupling) when $\Phi$ changes $\pi$. This is surprising, since conventional photon-mediated coupling theory predicts that $G \sim e^{i\Phi}$, changing from purely real to purely imaginary values when $\Phi$ changes $\pi/2$ \cite{van2013photon,tiranov2023collective}. The anomalous doubled phase period is also observed by setting the cavity near the 2nd critical coupling condition at $d$ = 5.87 mm \cite{Supp}.

\begin{figure} [!t]
\begin{center}
\includegraphics[width=8.7 cm]{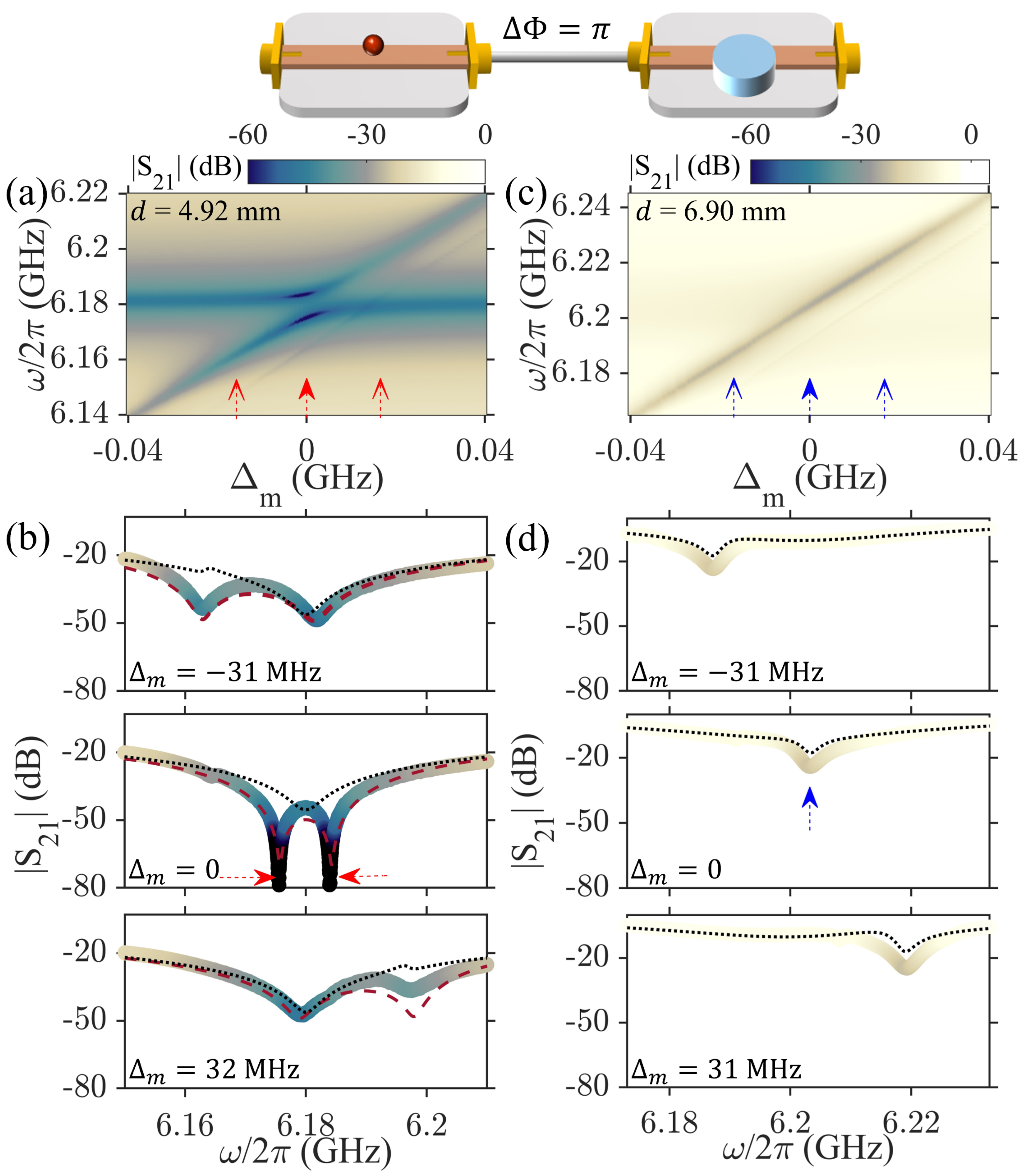}
\caption{\textbf{Long-distance coupling: the field dependency.} The left and the right panels compare the results measured at $d$=4.92 and 6.90 mm, respectively. (a) $\&$ (c) $|S_{21}|$ mapping plotted as a function of the probing frequency $\omega$ and the field detuning $\Delta_m$, showing level repulsion caused by remote coupling. (b) $\&$ (d) Transmission spectra plotted with the same colorbar as in the top panel. Hybridized modes are indicated in (b) by the red arrows at $\Delta_m = 0$.  The black dotted curves and the red dashed curves are theoretical calculations using Eq. \ref{eq:s21} with $\eta = 1, \delta = 1$ and $\eta = 2, \delta = 0.996$, respectively.}
\label{Fig4}
\end{center}
\end{figure}

In contrast, at $d$ = 6.90 mm, the result is different as shown in Figs. \ref{Fig3}(d)-(f). Here, no mode splitting is found, instead, a single resonance with a negative group delay $\tau_g$ is observed. A noteworthy feature is that $1/|S_{21}(\omega_c)|$ shows a phase-dependent oscillation, which we will analyze later by using the coupled mode theory.

To confirm these typical features, we perform the 2nd experiment at different field detuning $\Delta_m = [\omega_m(H)-\omega_c]/2\pi$, while fixing $\Delta\Phi = \pi$. Here, we compare the field dispersions measured at $d$ = 4.92 and 6.90 mm.

At $d$ = 4.92 mm, the measured dispersion plotted in Fig. \ref{Fig4}(a) reveals level repulsion between the remote magnon and cavity modes, and the spectra plotted in Fig. \ref{Fig4}(b) show anti-crossing with amplitudes exchange. This demonstrates coherent energy exchange between the two remote modes. From the splitting measured at $\Delta_m$ = 0, we determine the coupling rate $|G|$ = 4.18 MHz, corresponding to a cooperativity $C = |G|^2/|\beta\alpha|$ = 7.5, which confirms that the critically-driven cavity magnonics operates in the strong-coupling regime \cite{rameshti2022cavity}.  In contrast, at $d$ = 6.90 mm, neither level repulsion nor anti-crossing are observed. The magnon mode appears to be superimposed on the broad background of the cavity mode.

Thus, two sets of experiments consistently reach the same conclusion: by setting the cavity near the critical coupling condition, long-distance coherence is established between the remote magnon and photon modes, revealing an anomalous long-distance strong coupling mediated by travelling photons.

\textit{Phenomenological model.--}
Exploring beyond the simple model of Eq. \ref{eq:g}, we now quantitatively analyze the data using the coupled mode theory \cite{fan2003temporal}. As depicted in Fig. \ref{Fig1}(b), the magnon ($\hat{m}$) and cavity ($\hat{c}$) modes, with intrinsic damping rates $\alpha_0/2\pi$ and $\beta_0/2\pi$, respectively, are coupled to the travelling photons with rates $\kappa_{m,L(R)}/2\pi$ and $\kappa_{c,L(R)}/2\pi$, respectively. The remotely coupled system is driven by the input field $\hat{s}_{+1}$ and probed by the output field $\hat{s}_{-2}$. The coupled equations for $\hat{c}$ and $\hat{m}$ are:
\begin{equation}
\begin{split}
\begin{pmatrix} \dot{\hat{c}} \\ \dot{\hat{m}} \end{pmatrix}
=& - i \begin{pmatrix}
\omega_c - i\beta_0 - i\kappa_{c}   & -i e^{i\Phi/\eta} \sqrt{\kappa_{c,R} \kappa_{m,R}}  \\
-i e^{i\Phi/\eta} \sqrt{\kappa_{c,L} \kappa_{m,L}}   & \omega_m - i\alpha_0 - i\kappa_{m}
\end{pmatrix}
\begin{pmatrix} \hat{c} \\ \hat{m} \end{pmatrix} \\
&-i\begin{pmatrix}  e^{i\Phi/\eta}\sqrt{\kappa_{c,R}}  \\ \delta \sqrt{\kappa_{m,R}} \end{pmatrix} \hat{s}_{+1}, \\
\hat{s}_{-2} = ~& \hat{s}_{+1} - i\begin{pmatrix}   e^{-i\Phi/\eta}\sqrt{\kappa_{c,R}}~~ & \delta \sqrt{\kappa_{m,R}} \end{pmatrix}
\begin{pmatrix} \hat{c} \\ \hat{m} \end{pmatrix},
\end{split}
\label{eq:eom}
\end{equation}
where $\kappa_{c(m)} = [\kappa_{c(m),L}+\kappa_{c(m),R}]/2$. We introduce $\eta$ and $\delta$ as two phenomenological parameters to account for the anomalous period and coupling rate, respectively.

Using Eq. \ref{eq:eom}, the transmission $S_{21} = \hat{s}_{-2}/\hat{s}_{+1}$ of coupled system is derived as,
\begin{equation}
\begin{split}
S_{21}
=&\frac{[\omega-\tilde{\omega}_m +i(1-\delta^2)\kappa_{m,R}](\omega - \tilde{\omega}_c) - G_0^2}
{(\omega-\tilde{\omega}_m + i \kappa_{m,R})(\omega - \tilde{\omega}_c + i \kappa_{c,R}) +  K e^{i2\Phi/\eta} },\\
\end{split}
\label{eq:s21}
\end{equation}
where $G_0^2 = -\kappa_{c,R} \kappa_{m,R}(1-\delta)\left(e^{i2\Phi/\eta}\sqrt{\frac{\kappa_{c,L}\kappa_{m,L}}{\kappa_{c,R}\kappa_{m,R}}}-\delta\right)$ and $K=\sqrt{\kappa_{c,R}\kappa_{m,R}\kappa_{c,L} \kappa_{m,L}}$.

Setting $\eta = 1$ and $\delta = 1$, Eqs. \ref{eq:eom} and \ref{eq:s21} reproduce the standard theory describing photon-mediated coupling \cite{fan2003temporal}. Using the parameters calibrated in Table \ref{Table} for $d$ = 6.90 mm and setting $\Phi = (2n+1)\pi$ with $n$ as an integer \cite{Supp}, the calculated $|S_{21}(\omega)|$ spectra plotted by the dotted curves in Fig. \ref{Fig4}(d) qualitatively agree with the measured spectra. Furthermore, as shown by the dotted curve plotted in Fig. \ref{Fig3}(f), the $\Phi$-dependence of $1/|S_{21}(\omega_c)|$ calculated for $\Phi \in [2n\pi, 2(n+1)\pi] $ qualitatively agrees with the measured data \footnote{The parameters $\beta$ and $\kappa_c$ change slightly by changing $\Phi$, which we have calibrated in \cite{Supp}}, showing that the observed oscillation originates from the photon-mediated coupling term $K e^{i2\Phi}$ in Eq. \ref{eq:s21}. Smoking guns of deviation from the standard theory are evident, but hard to verify from the experiments performed at such non-critical coupling settings.

In contrast, at $d$ = 4.92 mm, $|S_{21}(\omega)|$ calculated from the standard theory ($\eta = 1$ and $\delta = 1$) fails completely to reproduce the observed splitting at $\Delta_m = 0$, as shown by the black dotted curve plotted in Fig. \ref{Fig4}(b). Moreover, it fails to explain the observed amplitude of the magnon-like mode measured at $\Delta_m \neq 0$. Here, we have to set $\delta = 0.996$ to reproduce the observed mode splitting at $\Delta_m = 0$, as shown by the red dashed curve plotted in Fig. \ref{Fig4}(b). Furthermore, we need to set $\eta = 2$ to account for the anomalous phase period, as shown by the red dashed curves plotted in Figs. \ref{Fig3}(a) and (b). Setting $\eta = 2$ also reproduces the amplitude of the magnon-like mode, as shown by the dashed curves plotted in Fig. \ref{Fig4}(b). Thus, critically-driven cavity magnonics reveals two anomalous features in contrast to conventional cases \cite{fan2003temporal,lalumiere2013input,buchmann2015quantum,karg2019remote}: (i) the photon-mediated coherence is no longer exactly erased by the photon-induced decoherence, and (ii) the phase period of the coupling strength is doubled.

Currently, there is no theory suitable for explaining these anomalies. To exam whether our discovery is limited to special cavities with specific geometric parameters, we replace the 3D cylindrical cavity with a 2D complementary split-ring resonator. Near the critically-driven condition, the experimental results \cite{Supp} well reproduce both anomalies, indicating that the phase-dependent anomalous coupling may be universal and deserves deep investigation.

\textit{Conclusion.--}
By coupling a YIG sphere remotely through cables over 2 meters long with either a cylindrical cavity or a ring resonator, we discover an anomalous long-distance strong coupling. Our experiments show that critically-driven cavity functions as a loaded high-Q oscillator with nearly zero damping, which is extremely sensitive to detect long-distance coupling. Phase-dependent measurements show unambiguously that the observed remote coupling is mediated by the travelling photons, and the field-dependent measurements directly demonstrates coherent energy exchange between a pair of local oscillators separated by a macroscopic distance. We find that the coupling strength oscillates exactly twice the period of conventional photon-mediated couplings. Both anomalies are found independent of specific geometric parameters of the cavity, demonstrating that critically-driven cavity magnonics is a robust platform for harnessing long-distance coherence, which may open new horizons for developing cavity magnonics network. The observed anomalies may intrigue broad interest for theoretical and experimental studies of photon-mediated coherence and remote sensing in coupled system of distributed oscillators.

This work has been funded by NSERC Discovery Grants and NSERC Discovery Accelerator Supplements (C.-M. H.). H.-K. L. and P.-T. F. acknowledges support from the Natural Sciences and Engineering Research Council of Canada (NSERC RGPIN-2021-02637) and Canada Research Chairs (CRC-2020-00134). Y.X. acknowledges support from the National Natural Science Foundation of China under Grant No. 61974067.

\end{document}